\documentclass[10pt,preprint,aps,amsmath,amssymb,prd]{revtex4}

\usepackage{graphicx}% Include figure files
\usepackage{bm}% bold math
\usepackage{color}

\begin{document}

\def\be{\begin{equation}}
\def\ee{\end{equation}}
\def\d{\mbox{\rm d}}
\def\LCDM{$\Lambda$CDM\,}

\title{Logical Entropy and Negative Probabilities in Quantum Mechanics}

\author{Giovanni Manfredi}
\email{giovanni.manfredi@ipcms.unistra.fr}
\affiliation{Universit\'e de Strasbourg, CNRS, Institut de Physique et Chimie des Mat\'eriaux de Strasbourg, UMR 7504, F-67000 Strasbourg, France}

\date{\today}

\begin{abstract}

The concept of Logical Entropy, $S_L = 1- \sum_{i=1}^n p_i^2$, where the $p_i$ are normalized probabilities, was introduced by David Ellerman  in a series of recent papers. Although the mathematical formula itself is not new, Ellerman provided a sound probabilistic interpretation of $S_L$ as a measure of the distinctions of a partition on a given set. The same formula comes across as a useful definition of entropy in quantum mechanics, where it is linked to the notion of purity of a quantum state.
The quadratic form of the logical entropy lends itself to a generalization of the probabilities that include negative values, an idea that goes back to Feynman and Wigner. Here, we analyze and reinterpret negative probabilities in the light of the concept of logical entropy. Several intriguing quantum-like  properties of the logical entropy are derived and discussed in finite dimensional spaces. For infinite-dimensional spaces (continuum), we show that, under the sole hypothesis that the logical entropy and the total probability are preserved in time, one obtains an evolution equation for the probability density that is basically identical to the quantum evolution of the Wigner function in phase space,  at least when one considers only the momentum variable.
This result suggest that the logical entropy plays a profound role in establishing the peculiar rules of quantum physics.
\end{abstract}

\maketitle

\section{Introduction}\label{sec:intro}
As its title suggests, this work sits at the crossroad of three different topics: (a) an alternative definition of entropy, (b) the extension of standard probabilities to negative values, and (c) the relevance of the first two items  to our understanding of quantum mechanics.
Here, we will introduce each topic separately, before bringing them together in the following sections.

\subsection{Logical Entropy}\label{sec:logicalentropy}
``Logical entropy" is a concept introduced by David Ellerman in a series of works spanning the last decade \cite{Ellerman2009,Ellerman2018}; see also Ellerman's  paper in this Special Issue. Succinctly, logical entropy is based on the concept of {\em distinctions}. If a certain set $U$ is partitioned into a number $n$ of subsets $B_i$ (such that $\cup_{i=1}^{n} B_i = U$), each endowed with a probability $p_i$ of finding an element of $U$ in that subset, then the probability that in two independent draws one will obtain elements in distinct subsets $B_i$ and $B_{j \neq i}$ is: $p_i(1-p_i)$. This is precisely the concept of {\em distinction}, i.e., the ability to establish that two independent  draws are different from one another.

Summing over all $n$ subsets, we obtain the total number of distinctions, which is the definition of the logical entropy $S_L$:
\be
S_L= \sum_{i=1}^n p_i(1-p_i) = 1- \sum_{i=1}^n  p_i^2 ,
\label{eq:def-entropy}
\ee
where we used the fact that $\sum_i p_i =1$. The subsets $B_i$ can possibly contain one single element, in which case $S_L$ represents the probability that two consecutive draws yield different elements of $U$. In this work, we will mainly consider this case, unless otherwise stated.
It is clear that $0 \le S_L \le 1$. The lower bound is reached when  one element has probability $p_i=1$, while for all others $p_{j \neq i}=0$. For equal probabilities ($p_i={1\over n}, \, \forall i$), one gets: $S_L=1-{1\over n} \to 1$, when $n\to \infty$.

Following \cite{Brukner1999,Manfredi2000}, one can also define the information $I$ as the complement of the entropy to unity:
\be
I = 1-S_L = \sum_{i=1}^n  p_i^2.
\label{eq:def-information}
\ee
This quantity reflects the knowledge we have of the state of a physical system, being maximum when we know its state with certainty, and minimum when all states are equally probable
\footnote{As an aside, we note that the idea of information as distinctions (differences, distinguishability, and diversity) would take the higher logical entropy states as making more distinctions or showing more diversity and distinguishability between the outcomes.  In that sense, higher logical entropy states may be thought as having {\em more}, rather than less, information. But here we stick to the definition of information as presented in the main text, which is the way it is usually interpreted in physics.}.
The information $I$ has the nice property of being the square of a norm in ${\mathbb R}^n$, actually the Euclidean norm.
This connection to Euclidean geometry allows one to use standard geometrical concepts when making use of the logical entropy. For instance, one can define the the scalar product: $p \cdot q = \sum_{i=1}^n  p_i q_i$ between two probability distributions $\{p_i\}$, $\{q_i\}$, and their Euclidean distance $d(p,q)$   as:
\be
d^2(p,q) = \sum_{i=1}^n(p_i-q_i)^2.
\label{eq:def-distance}
\ee

Of course, the logical entropy definition \eqref{eq:def-entropy} implies very different properties from the standard Shannon-Von Neumann entropy
\be
S_{VN} = -\sum_{i=1}^n p_i \log p_i .
\ee
In particular $S_{VN} $ is additive, while $S_L$ is not, at least not in the standard fashion, see \cite{Wehrl1978,Manfredi2000}. For a system known with certainty, both entropies yield, $S_{VN}=S_L=0$, but for maximal uncertainty $S_{VN}=\log n$, whereas $S_L=1-{1 \over n}$.

Again we emphasize that, in contrast to the  Shannon-Von Neumann entropy, the logical entropy  $S_L$ represents both a {\em probability} (of obtaining different results in two consecutive draws, as mentioned above) and a {\em norm} in the Euclidean space ${\mathbb R}^n$. These facts have important consequences, as we will see in the next section.

Although Ellerman \cite{Ellerman2009,Ellerman2018} provided a solid and fruitful probabilistic interpretation of this definition of entropy, the formulae \eqref{eq:def-entropy} and \eqref{eq:def-information} are not new. Quite the contrary, they have been discovered and rediscovered many times in the past, in very different areas of research.  In biology and ecology, $S_L$ is known as the Gini-Simpson index \cite{Simpson1949,Hunter1988,Crupi2019}, which quantifies the diversity of species in an ecosystem.
It was used by Polish mathematicians (and then by Alan Turing himself) to find patterns in messages generated by the Enigma machine during World War 2 \cite{Christensen2007}.
In statistical mechanics, $S_L$ is a special case of the Tsallis entropy \cite{Tsallis1988} with index $q=2$.
In quantum physics, a version of $S_L$ was used to quantify our knowledge of the state of a quantum system \cite{Brukner1999,Brukner2003}. It was also shown to be particularly adapted to the Wigner phase-space representation of quantum mechanics \cite{Manfredi2000}.

\subsection{Negative Probabilities}\label{sec:negprob}
The very definitions of $S_L$ and $I$ lend themselves to the natural generalization whereby the probabilities $p_i$ can take negative values. This is in analogy with vectors in ${\mathbb R}^n$, which can indeed have negative components, although their norm remains positive.

Negative probabilities have a long history of interest, especially among physicists struggling to make sense of some of the weird properties of quantum mechanics. Feynman \cite{Feynman1987} was one of the first to ponder the meaning of negative probabilities in a quantum context (although he published his ideas in 1987 in a volume in honor of David Bohm, he states there that he developed these reflections some twenty years earlier). For Feynman, negative probabilities should be considered as a useful bookkeeping tool just like negative numbers
\footnote{The need for negative numbers can be circumvented through the trick of double-entry bookkeeping, see \cite{Ellerman1985}.}. As an example, he mentions a man starting a day with five apples, giving away ten at midday and earning eight in the evening. The initial (5) and final (3) numbers of apples owned by the man are both positive and thus unambiguous to interpret. But if we take the numbers at face value, the man will have $-5$ apples some time in the afternoon, which does not quite make sense unless we postulate that one  is allowed to count the number of apples only in the morning and in the evening, but not in the middle of the day. Hence, negative probabilities are allowed as long as they intervene in contexts where they cannot be observed directly.
All this is reminiscent of the limitations on measuring some quantities, which are intrinsic to quantum physics \cite{Scully1994,Curtright2001}.

Of course, negative probabilities had appeared in quantum mechanics even earlier, when Wigner \cite{Wigner1932} introduced his celebrated pseudo-probability distribution in the classical phase space (``Wigner function"), which almost always takes negative values. Indeed, the negativity of a Wigner function can be used as a tool to quantify the degree of quantumness of a particular state, as was done even experimentally \cite{Deleglise2008}.

Negative probabilities have also been studied in a fundamental mathematical context \cite{Bartlett1945,Khrennikov2008,Khrennikov2009,Burgin2010} and for applications to financial modeling \cite{Burgin2012}.
A thorough, if not very recent, review on the topic of negative probabilities in physics was published in 1986 \cite{Muckenheim1986}, and contains quotations from several eminent scientists on this somewhat controversioal problem.

\subsection{Quantum Mechanics}\label{sec:quantum-mech}
The earliest relationship between negative probabilities and quantum mechanics dates back to Wigner \cite{Wigner1932}, who in 1932 introduced a pseudo-probability distribution in the phase space $(x,p)$ which possesses many of the properties of classical probability distributions (for instance, it can be used to compute averages using the classical formula), {\em except non-negativity}. The Wigner function $w(x,p,t)$ can describe both pure and mixed quantum states and evolves in time according to an integro-differential equation similar to the classical Liouville equation. Wigner functions have proven exceedingly useful in a variety of domains, ranging from condensed matter and nanophysics, to quantum plasmas and quantum optics (see \cite{Hillery1984} for a review).

The Wigner equation conserves in time not only the total probability $\int\int w(x,p,t) dx dp$, but also the integral of the square of the Wigner function: $\int\int w^2(x,p,t) dx dp$. Note that higher powers $\int\int w^r dx dp$, with $r>2$, are not conserved, in contrast to the classical Liouville equation, for which the conservation property is valid for any value of $r$.
Some time ago, the present author suggested that one uses
\be
S_L = 1-I = 1- h \int\int w^2\,dx \,dp
\label{eq:def-entropy-wigner}
\ee
as the definitions of entropy and information \cite{Manfredi2000}, where $h$ is Planck's constant (this is necessary to render the integral term in the above expression non-dimensional). Equation \eqref{eq:def-entropy-wigner} can be viewed as the continuous counterpart of Eq. \eqref{eq:def-entropy}, i.e., its extension to an infinite dimensional space. Also note that the logical entropy can be expressed in terms of the trace of the density operator, as  $S_L=1 - \rm Tr(\hat \rho^2)$.

 More recently, negative probabilities have been explored in various quantum mechanical contexts, such as indistiguishability \cite{deBarros2020}, quantum computation \cite{Veitch2012}, and contextuality \cite{Spekkens2008}.
Besides, an operational interpretation of negative probabilities has been proposed by Abramsky and Brandenburger \cite{Abramsky,Abramsky2014}. In \cite{Abramsky2014}, they propose a simple scenario to illustrate pedagogically the use of negative probabilities in quantum mechanics, by considering a system comprising two bit registers.

The rest of this work is devoted to the study of the properties of the logical entropy \eqref{eq:def-entropy} and information \eqref{eq:def-information} when one relaxes the requirement that $p_i \ge 0,\, \forall i$. It will be claimed that the logical entropy constitutes the natural framework for the introduction of negative probabilities. Interestingly, by combining the definition of logical entropy with negative probabilities, one can recover many properties that are typical of quantum systems.

The main result obtained here is that, simply by requiring the logical entropy to be conserved in time, one obtains an evolution equation for the probability density that is virtually identical to the evolution equation of the Wigner function in physics,  at least when one considers only the momentum variable.
This remarkable result suggest that the logical entropy plays a profound role in establishing the peculiar rules of quantum physics.

\section{Finite-dimensional spaces}\label{sec:finite-dim}

We consider a set of $n$ outcomes, each endowed with probability $p_i$. The probabilities satisfy
\begin{eqnarray}
\sum_{i=1}^n p_i &=& 1, \label{eq:sum-p}\\
\sum_{i=1}^n p_i^2 &=& R^2, \label{eq:sum-p2}
\end{eqnarray}
where $0\le R \le 1$. Then the logical entropy and information are, respectively, $S_L=1-R^2$ and $I=R^2$. The number $R$ can be interpreted as the Euclidean norm of the vector $p=(p_1, \dots p_n)$ in ${\mathbb R}^n$: $\|p\| =R$.
Geometrically, Eqs. \eqref{eq:sum-p} and \eqref{eq:sum-p2} represent respectively a hyperplane and a hypersphere of radius $R$ in ${\mathbb R}^n$, and their intersection yields the probability distributions $\{p_i\}$ satisfying those equations.

In analogy with quantum physics, we shall call {\em pure states} the probability distributions for which $R=1$ (corresponding to maximum information and minimum entropy) and {\em mixed states} those for which $R<1$.
Indeed, Wigner functions for pure and mixed quantum states satisfy precisely these properties, when the entropy is defined as in Eq. \eqref{eq:def-entropy-wigner}.
If we request all probabilities to be nonnegative, then the only pure states are those for which $p_i=1$ and $p_{j\neq i}=0$, that is, the $i$-th outcome  can be predicted with certainty.
However, if we admit negative probabilities, there exist other pure states with some $p_i<0$ which still satisfy Eqs. \eqref{eq:sum-p}-\eqref{eq:sum-p2} with $R=1$.

 To dissipate all ambiguities, here we are not dealing with ``probability amplitudes" as in quantum mechanics. Probability amplitudes are complex quantities, while our $p_i$s are real numbers, albeit potentially negative. Our approach is the same as the one based on Wigner functions (also real quantities), which represent quantum states with real, but signed, numbers. 

In the rest of the present section, we will focus on the cases $n=2$, which is trivial and does not admit negative probabilities, and $n=3$, which is much richer.
The infinite-dimensional case will be treated in Sec. \ref{sec:infinite-dim}.

\subsection{General properties for $n=2$ and $n=3$}

\begin{figure}[]
\centering
\includegraphics[width=0.37\linewidth]{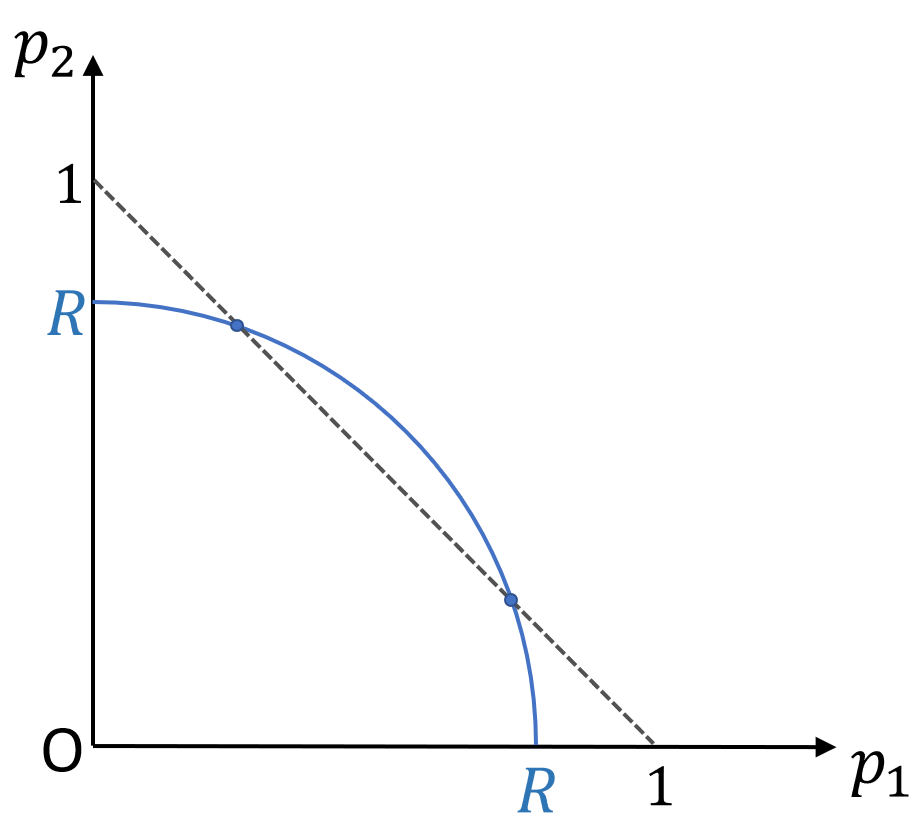}
    \caption{Schematic representation of the case $n=2$. The total probability constraint \eqref{eq:sum-p} is represented by the dashed straight line, while the entropy constraint \eqref{eq:sum-p2} is represented by the blue quarter circle of radius $R$. Solutions are given by their intersections.}
    \label{fig:1}
\end{figure}

For $n=2$, the solution is given by the intersection of the straight line and the circle shown in Fig. \ref{fig:1}. It is clear that for $R \le 1$, only positive solutions are allowed. Solving Eqs. \eqref{eq:sum-p}-\eqref{eq:sum-p2} yields $p_{1,2} = \frac{1\pm \sqrt{2R^2-1}}{2}$.
No solutions exist for $R< \sqrt{2}/2$ (dashed straight line tangent to the circle). For this value of $R$, one obtains $p_1=p_2=1/2$, which is the maximally mixed state (with largest entropy $S_L=1/2$).
%The fact that only positive probabilities are allowed for the $n=2$ case should not come as a surprise: it corresponds, in quantum mechanics, to the fact that a two-level system (qubit) can be represented classically by two angles on the Bloch sphere.}}

\begin{figure}[]
\centering
\includegraphics[width=0.45\linewidth]{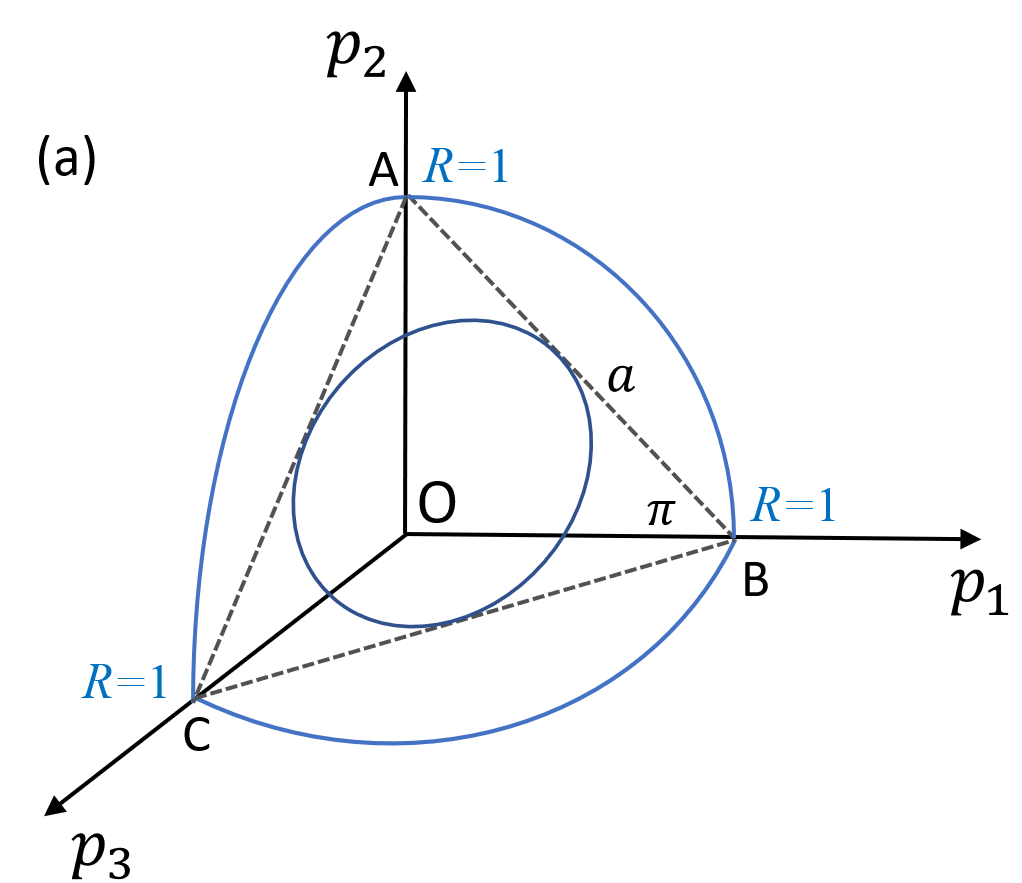}
\includegraphics[width=0.35\linewidth]{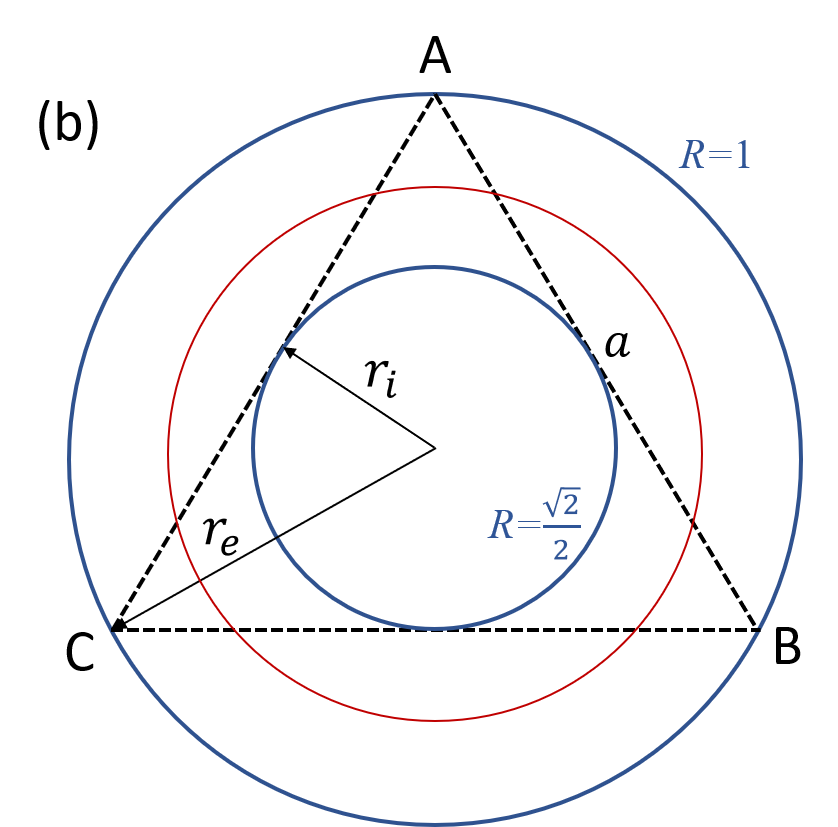}
\includegraphics[width=0.25\linewidth]{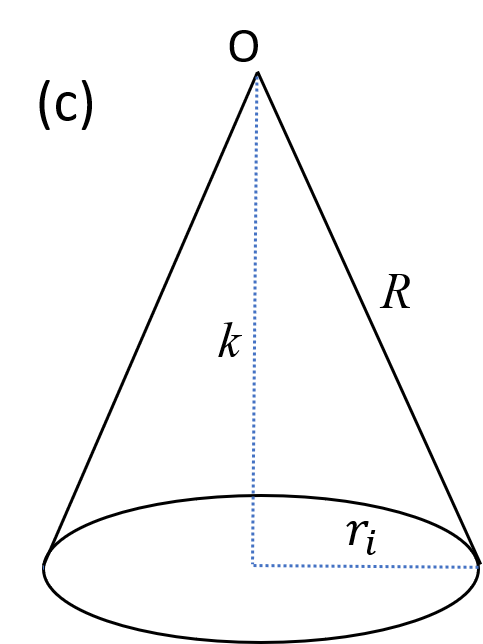}
    \caption{Schematic representation of the case $n=3$. (a) The solutions of Eqs. \eqref{eq:sum-p}-\eqref{eq:sum-p2}  lie on the circle given by the intersection of the sphere of radius $R$ (here represented for the pure state with $R=1$) and the plane $\pi$ passing through the points A, B and C. The dashed triangle ABC, lying on the plane $\pi$, has all sides equal to $a=\sqrt{2}$. The inner circle is inscribed into the triangle. (b) View of the plane $\pi$ with the circumscribed and inscribed circles of radii  $r_e$ and $r_i$, which correspond, respectively, to values $R=1$ and $R=1/\sqrt{2}$. The thin red circle is an intermediate case where positive and negative probabilities coexist (the latter lie outside the triangle). (c) Circular cone with apex at the origin $O$ and basis circle of radius $r_i$. The height $k$ of the cone is the distance between the origin and the plane $\pi$.
    }
    \label{fig:2}
\end{figure}

The case $n=3$ is depicted schematically in Fig. \ref{fig:2}(a) for the special case $R=1$ (pure states). It is evident that there are three pure states with nonnegative probabilities: $(1,0,0)$, $(0,1,0)$ and $(0,0,1)$, which represent certainty for one of the three possible outcomes. These states form an orthonormal basis which we denote by ${\textbf e}_i$.
However, there exist an infinity of other pure states with negative probabilities. These are the states that lie on the circle given by the intersection between the sphere of radius $R$ and the plane $\pi$ defined by the three vectors  ${\textbf e}_i$. Actually, \emph{all} pure states, except ${\textbf e}_1$, ${\textbf e}_2$ and ${\textbf e}_3$, feature some negative probabilities. A simple example is the state: $p=(2/3, 2/3, -1/3)$.

A view of the plane $\pi$ is shown in Fig. \ref{fig:2}(b). The circles represent the intersections of the plane and the sphere, for different values of the radius $R$. Points that lie outside the equilateral triangle ABC (with sides $a=\sqrt{2}$) have negative probabilities. The circumscribed circle, corresponding to $R=1$, has radius $r_e=a/\sqrt{3}=\sqrt{2/3}$.

For information $I=R^2$ smaller than unity, i.e. for mixed states, there are some positive and negative solutions (thin red circle in Fig. \ref{fig:2}(b)). Further decreasing $R$, we reach the situation of the inner circle of radius $r_i=r_e/2=\sqrt{6}/6$, for which all probabilities are positive. To determine the value of $R$ corresponding to $r_i$, we consider the cone of vertex $O$ and base radius $r_i$, see Fig. \ref{fig:2}(c). The height $k$ of the cone is the distance between the origin $O$ and the plane $\pi$, which turns out to be $k=1/\sqrt{3}$.
From this, we deduce that the radius $R$ corresponding to the inner circle in Fig. \ref{fig:2}(b) is $R=1/\sqrt{2}$.
Finally, for $R=k=1/\sqrt{3}$,  the sphere is tangent to the plane $\pi$, and the only solution is $p_1=p_2=p_3=1/3$, corresponding to maximum entropy $S_L=2/3$. For smaller $R$, there are no solutions.

In summary, defining the radii $R_{max}=1$, $R_{pos}=1/\sqrt{2}$, and  $R_{min}=1/\sqrt{3}$, we obtain that:
\begin{itemize}
\item
For $R_{pos} < R \le R_{max}$, there exist some negative-probability solutions;
\item
For $R_{min} \le R \le R_{pos}$, there exist only positive-probability solutions;
\item
For $R = R_{min}$: maximum entropy solution $p_1=p_2=p_3=1/3$;
\item
For $R < R_{min}$, there exist no solutions.
\end{itemize}
The above considerations can be easily extended to $n>3$, yielding: $R_{max}=1$, $R_{pos}=1/\sqrt{n-1}$, and  $R_{min}=1/\sqrt{n}$, with maximum entropy solution: $p_i=1/n, \, \forall i$. We note that for large $n$, one has $R_{min} \approx R_{pos} \sim 1/\sqrt{n}$. Therefore, almost all existing solutions will display some negative values.

From Fig.  \ref{fig:2}, it is evident that, for pure states, the most negative value of $p_i$ is reached when two probabilities are identical and positive, and the third one is negative, i.e., $p_1=p_2 =p$ and $p_3=-q$. Direct computation yields the result $p=2/3$ and $q=1/3$. The three vectors: ${\textbf u}_1 \equiv (2/3,2/3,-1/3)$, ${\textbf u}_2 \equiv (2/3,-1/3,2/3)$ and ${\textbf u}_3 \equiv(-1/3,2/3,2/3)$ also constitute an orthonormal basis in ${\mathbb R}^3$.

This reasoning can be extended to $n$ dimensions, yielding $p_1= \dots =p_{n-1} = {2 \over n}$ and $p_n = {{2-n} \over n}$. From this, one can construct an orthonormal basis $\{{\textbf u}_1 \dots {\textbf u}_n \}$. For instance, for $n=4$ one gets $p_1=p_2 =p_3=1/2$ and $p_4 = -1/2$.

Finally, we stress that, from this simple example with $n=3$, negative probabilities arise very naturally if the $p_i$ are requested to satisfy the two equations \eqref{eq:sum-p} and \eqref{eq:sum-p2}, which fix the total probability and total entropy (or information) of the system. Indeed, for a mixed state such as described by the thin red circle in Fig.  \ref{fig:2}(b), it would be odd to retain only the positive-probability solutions (inside the triangle) and discard the negative ones (outside the triangle). Hence, the entropy definition \eqref{eq:sum-p2} calls for the acceptance of negative probabilities on the same footing as positive ones.

\subsection{Maximization with constraints}
We would like to maximize the entropy $S_L$ (minimize the information) with a constraint. This is analogous to the statistical mechanics problem of finding the equilibrium probability distribution that maximizes entropy for given energy, which yields the Maxwellian distribution if one uses the Shannon-Von Neumann entropy.
Let us call $X$ our constraint, which has the mean value $m \equiv \langle X \rangle = \sum_i p_i X_i$. The functional $F$ to be minimized is given by the information $I$ augmented by two constraints on the total probability and  the average of $X$:
\be
F = \sum_i p_i^2 - \lambda \sum_i p_i + \mu \sum_i p_i X_i,
\label{eq:functional-F}
\ee
where $\lambda$ and $\mu$ are Lagrange multipliers. Setting the variation of $F$ to zero, i.e.:
\[
\delta F = 2\sum_i p_i \delta p_i - \lambda \sum_i \delta p_i  + \mu \sum_i X_i \delta p_i =0,
\]
one gets
\be
p_i = \frac{\lambda -\mu X_i}{2}.
\label{eq:equilibrium}
\ee
The Lagrange multipliers are determined by using the constraints: $\sum_i p_i =1$ and $\sum_i p_i X_i = m$.

As an example, we take again $n=3$ and $X_i = (-1,0,1)$. This choice yields $\lambda=2/3$, $\mu=-m$, and the ``equilibrium" probability distribution:
\be
p = \left( {1 \over 3} -{m \over 2}, \, {1 \over 3}, \, {1 \over 3} +{m \over 2} \right).
\ee
The total information is $I=R^2 = {1 \over 3} +{m^2 \over 2}$. As it must be smaller or equal to unity, we have a constraint on the maximum mean value allowed for the variable $X$: $m \le {2\over \sqrt{3}} \equiv m_{\rm max} \approx 1.15$. For $m=m_{\rm max}$, we obtain the pure state
\be
p = \left( {{1-\sqrt{3}} \over 3} , \, {1 \over 3}, \, {{1+\sqrt{3}} \over 3} \right),
\label{eq:maxuneven}
\ee
for which $p_1 <0$.
Indeed, $p_1$ is negative whenever $2/3<m<m_{\rm max}$.
 For $m<2/3$ all probabilities become positive and for $m=0$ we recover the maximally mixed state with all probabilities equal to 1/3.
 Similar considerations apply for the symmetric cases with negative $m$.

 The above situation can be viewed as that of a die with three faces. For $m=0$ the die is even, and all faces are equally probable. Hence, $|m|$ may be interpreted as an index of unevenness of the die. Classically, i.e. only allowing positive probabilities, the most uneven die is obtained for $m=2/3$, yielding the state $p=(0,1/3,2/3)$ (for $m=-2/3$ the roles of $p_1$ and $p_3$ are interchanged), which has information $I=5/9$.
 But if we admit negative probabilities, $m$ can be increased up to $m_{\rm max}={2\over \sqrt{3}}$, which gives the state of Eq. \eqref{eq:maxuneven}, with information $I=1$.

 \subsection{Interpretation}
 The existence of negative probabilities induces some nonstandard properties that are reminiscent of the paradoxes encountered in quantum physics.
 For example, let us consider a pure state $p$ with $n=3$ and assimilate the three possible outcomes to the colors of marbles drawn from a bag: red (R), blue (B) and green (G). Like for all pure states, the probability to get the same color in two consecutive draws is $I=1$, while the probability to get different colors is $S_L=0$. Let us suppose that we draw a number marbles, but do not look at their colors for the moment [Fig. 3(a)]. Then we look at the second and third marble and observe that they have the same color (as they should), namely red. Subsequently, we look at the sixth and seventh marble and notice they are both blue [Fig. 3(b)].

 So far, all is in agreement with our expectations. But what would have happened if we had first looked at marbles number 3 and 6 [Fig. 3(c)]? According to the previous ``experiment", they should be of different colors (red and blue), but this is not allowed by the probability distribution of a pure state. Hence, we should find that they have the same color, which is in contradiction with the experiment (b) on the figure.
 We are forced to conclude that the marbles do not have a predefined color prior to the observation, something that is typical for quantum objects \cite{Bell1966,Kochen1968,Kochen1975}.

 As a second example, let us consider two probability distributions $p=({2 \over 3},{2 \over 3},-{1 \over 3})$ and $q=(-{1 \over 3},{2 \over 3},{2 \over 3})$, which we can be visualized as two different bags containing, respectively, red (R), blue (B) and green (G) marbles in different proportions. They are both pure states and orthogonal to each other, $p \cdot q = \sum_i p_i q_i=0$. The latter property means that  the outcomes of the two bags are perfectly anticorrelated, i.e. if the outcome of the first bag is R then that of the second bag must be not~R (denoted $\overline{\rm R}$). We draw pairs of marbles from each bag. From the second bag, the probability of drawing a pair of red marbles is: $\rm Prob_q(RR) = q_1^2 = \frac{1}{9}$. Since the outcome of the first bag is perfectly anticorrelated with that of the second bag, this number should also represent the probability of {\em not} drawing a pair of red marbles from the first bag. However, if we compute the same probability using the distribution $p$ of the first bag, we obtain: $\rm Prob_p(\overline{\rm R}\overline{\rm R}) = Prob_p(BB)+ Prob_p(GG) = p_2^2 + p_3^2 = \frac{5}{9} $, which is manifestly different.

 This example shows that the following two procedures are mutually exclusive: (i) drawing one marble from bag 1 and another from bag 2, which gives perfectly anticorrelated results; (ii) drawing two marbles from either bag, which yields perfectly correlated results.
 If two experimentalists draw a marble from each bag and then communicate their results, they always observe anticorrelation. However, once they have done so, they cannot use this knowledge to predict their next draw by using the correlation property of each bag, because the latter is valid only if pairs of marbles are observed together. (Remember that the logical entropy quantifies {\em distinctions} between two draws, but says nothing about single draws. Indeed the outcome of a single draw is meaningless, as its probability can be negative; only pairs of consecutive draws are meaningful.)
 Similarly, if one experimenter observes BB in one bag and communicates this result to the second experimentalists, the latter cannot use it to predict that her next draw will be $\overline{\rm B}\overline{\rm B}$, because the anticorrelation property holds only as long as both elements of the draw are still unknown.
%These features are also reminiscent of the ``complementarity" properties of quantum observations.

\begin{figure}[]
\centering
\includegraphics[width=0.6\linewidth]{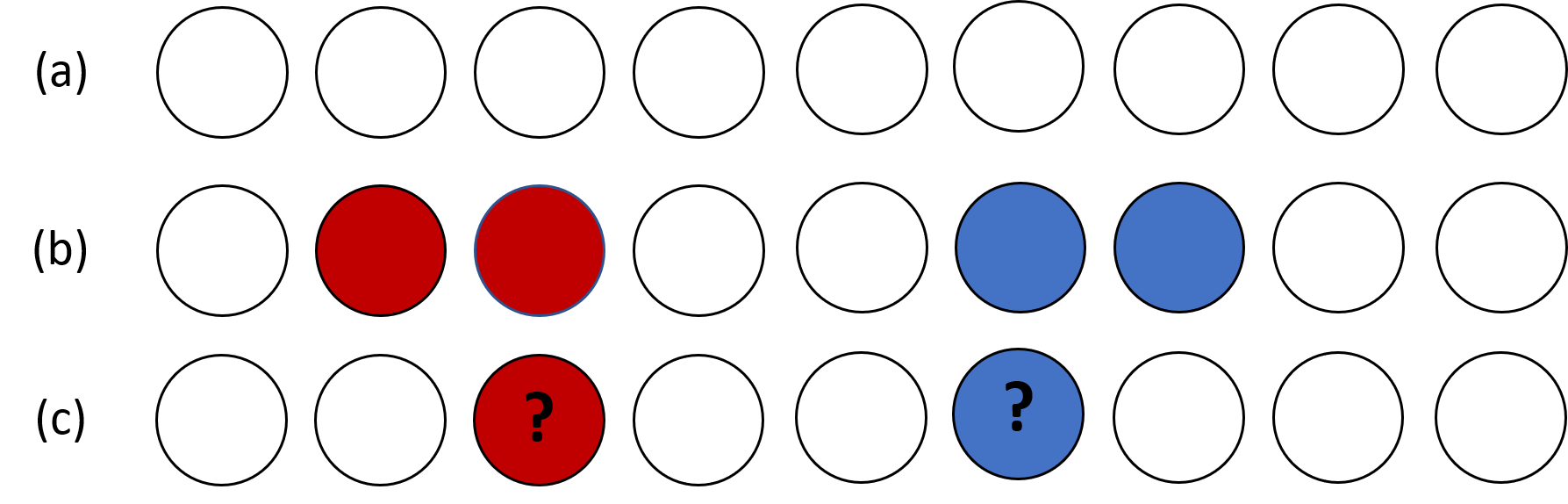}
    \caption{(a) Nine particles are drawn from a probability distribution $p$, corresponding to a pure state, but they are not yet observed. (b) We look at particles 2-3, which turn out to be both red, and then look at particles 6-7, which turn out to be both blue. (c) Had we drawn particles 3-6, we would have expected them to be of same color, but this is in contradiction with the ``experiment" of row (b).}
    \label{fig:3}
\end{figure}

 \subsection{Dynamics}
The probability distribution $p(t)$ should evolve in a way that preserves both the total probability (of course) and the total information or entropy. In 3D this is possible only if the vector $p$ performs a rotation around the axis perpendicular to the plane $\pi$ and going through the origin O, see Fig. \ref{fig:2}(a). This can be viewed as a rotation around the vector $v=(1/\sqrt{3},1/\sqrt{3},1/\sqrt{3})$, which yields the evolution equation
\[
\frac{dp}{dt} = v \times p(t),
\]
where $\times$ denotes the standard 3D cross product.
However, the representation using the vector product cannot be readily extended to dimensions $n>3$, so it is more useful to write the above equation in matrix form:
\be
\frac{dp_i}{dt} =   \frac{\sqrt{3}}{3} \, \sum_{j=1}^n M_{ij} \, p_j ,
\label{eq:evolut-3d}
\ee
where $\textsf{M} = \{ M_{ij}\}$ is the antisymmetric matrix
\be
\textsf{M} =
\begin{pmatrix}
0 & -1 & +1 \\
+1 & 0 & -1  \\
-1 & +1 & 0 \,,
\end{pmatrix}
\label{eq:evolut-matrix}
\ee
satisfying $M_{ij} = -M_{ji}$ and $\sum_i M_{ij} =\sum_j M_{ij} = 0$. The latter conditions guarantee that the total probability and the total information are indeed conserved during the evolution.

The above matrix form of the evolution equation \eqref{eq:evolut-matrix} is readily adapted to higher dimensions, and will be  generalized to infinite dimensional systems (continuum) in the next section.

\section{Infinite-dimensional spaces (continuum)}
\label{sec:infinite-dim}

\subsection{Generalities}
The logical entropy and information can be generalized to an infinite-dimensional system, i.e. in the continuum. We define the probability density $f(z)$, with $z \in \mathbb{R}$, normalized so that $\int_{-\infty}^{\infty}  f(z)\,dz=1$. Then the logical entropy and the information are defined as follows \cite{Manfredi2000}:
\be
S_L = 1-I = 1 - h \int_{-\infty}^{\infty}  f^2(z)\,dz
\label{eq:enbtropy-def-continuum}
\ee
where the constant $h$ has the same dimensions as $z$, and $f$ has the dimensions of  $h^{-1}$. The so-defined information is basically the $L^2$ norm in the space of real square-integrable functions.

Given the arbitrariness of the constant $h$, it is not automatic that $0 \le S_L \le 1$: some very peaked functions of $z$ may yield an entropy that is negative, or equivalently an information greater than unity.
Hence, we {\em require} that $0 \le S_L \le 1$, and restrict the space of allowed probability densities to those whose entropy satisfies this condition.

A useful bound on $f(z)$, which is reminiscent of the bound on Wigner functions \cite{Hillery1984}, can be obtained as follows. Let us consider pure states ($I=1$) and write
\[
h \int f^2(z) dz = 1 = \left( \int f(z) dz \right)^2 = \int\int f(x)f(y) dx dy =  \int\int f\left(z-{\lambda \over 2}\right)f\left(z+{\lambda \over 2}\right) dz d\lambda.
\]
This can be reformulated as
\[
\int dz\left[h f^2(z) - \int f\left(z-{\lambda \over 2}\right)f\left(z+{\lambda \over 2}\right) d\lambda \right] = 0.
\]
Setting the integrand equal to zero yields:
\[
hf^2(z) = \int f\left(z-{\lambda \over 2}\right)f\left(z+{\lambda \over 2}\right) d\lambda ,
\]
which is an integral equation for $f(z)$. Finally, using the Cauchy-Schwartz inequality, we get
\[
h f^2 \le
\left(\int  \left| f\left(z-{\lambda \over 2}\right)\right|^2 d \lambda \right)^{1/2}  \left(\int \left| f\left(z+{\lambda \over 2}\right)\right|^2 d \lambda \right)^{1/2}
= 2 \int f^2(z) dz = {2 \over h},
\]
from which we deduce the bound
\be
\max_z |f(z)| = \frac{\sqrt{2}}{h} .
\label{eq:f-bound}
\ee
Obviously, the above bound limits the peakedness of $f(z)$ for a given value of $h$.
For a mixed state with information $I<1$, the bound becomes: $\max |f| = \sqrt{2 I}/h$.

For instance, if the probability density is a Gaussian with standard deviation $\sigma$: $f(z)=e^{-z^2/2\sigma^2}/(\sqrt{2\pi}\sigma)$ and we require that $I=1$, we obtain
\[
\sigma = \frac{h}{2\sqrt{\pi}}.
\]
This value yields exactly the maximum of Eq. \eqref{eq:f-bound}, showing that the bound is saturated for a Gaussian distribution of unit information (pure state). For $\sigma > \frac{h}{2\sqrt{\pi}}$, we have $I<1$, i.e. a mixed state.

All this is similar to a bound that can be obtained on the quantum Wigner function $w(x,p)$ \cite{Hillery1984}, where $x$ and $p$ are respectively position and momentum: $\max_{x,p} |w(x,p)| = 2/h$, where here $h$ is Planck's constant. The additional factor $\sqrt{2}$ is due to the fact that the maximization is done in the 2D phase space $(x,p)$ instead of the 1D space $(z)$ considered above. These considerations establish a suggestive link between the present results and the properties of quantum mechanics, on which we will further elaborate in the forthcoming subsections.

\subsection{Dynamics}
 The time evolution of the probability density $f(z,t)$ must preserve both the total probability and the entropy, hence it has to be a rotation in the appropriate functional space. In analogy with the finite-dimensional case, see Eqs. \eqref{eq:evolut-3d} and \eqref{eq:evolut-matrix}, we write the general evolution equation for $f(z,t)$ as
 \be
 \frac{\partial f}{\partial t} = {1 \over h} \, \int M(z,z') f(z',t) \,dz' ,
 \label{eq:evolut-f}
 \ee
 where $M$ must be antisymmetric: $M(z,z') = -M(z',z$). In order to preserve the total probability in time, one should also have: $\int M(z,z')dz=0=\int M(z,z')dz'$, which follows immediately upon integrating \eqref{eq:evolut-f} over $z$.
 Further, by multiplying Eq. \eqref{eq:evolut-f} by $f(z,t)$ and integrating, we obtain
 \[
  \frac{d}{dt} \int f^2(z,t) dz= {2 \over h} \, \int \int f(z,t)M(z,z') f(z',t) dz' dz = 0.
 \]
The last equality follows because the function $\varphi(z,z') \equiv f(z,t)M(z,z') f(z',t)$ is such that $\varphi(z,z') = -\varphi(z',z)$, hence it is odd with respect to the diagonal of the $(z,z')$ plane, and integration over all such planes yields zero.

The two-variable function $M(z,z')$ can be conveniently written as $M(z,z') = m(z-z')$, where $m(\zeta)$ is a single-variable odd function: $ m(\zeta) = -m(-\zeta)$. The so-constructed $M(z,z')$ satisfies all the properties mentioned in the preceding paragraph. Hence, we rewrite:
\be
 \frac{\partial f}{\partial t} = {1 \over h} \, \int m(z-z') f(z',t) \,dz' ,
 \label{eq:evolut-f-mu}
 \ee
 We have included explicitly the constant $h$ in the evolution equation for further comparison with Wigner functions. With this choice, $m$ has the dimensions of an inverse time.

 We now write $m(\zeta)$ in terms of its Fourier transform
 \be
  m(\zeta) = i \int d\lambda \, \hat m(\lambda) \exp{\left(\frac{2\pi i \zeta \lambda}{h}\right)}.
 \label{eq:mu}
\ee
If $\hat m(\lambda) = -\hat m(-\lambda)$ then it follows that $m(\zeta)$ is indeed an odd function. As the Fourier transform of an odd real function is purely imaginary, we also have that $m(\zeta)$ is real, as intended. Note that $\lambda$ is dimensionless.
Let us now write the odd function $\hat m(\lambda)$ as follows, without loss of generality:
\[
\hat m(\lambda) = \Omega\left(a + {\lambda \over 2} \right) - \Omega\left(a - {\lambda \over 2} \right),
\]
where $a$ is a constant.
Inserting all these definitions into the evolution equation \eqref{eq:evolut-f-mu}, one gets
\be
 \frac{\partial f}{\partial t} = {i \over h} \, \int \int \left[ \Omega\left(a + {\lambda \over 2} \right) - \Omega\left(a - {\lambda \over 2} \right) \right] \exp\left(\frac{2 \pi i (z-z') \lambda}{h}\right) f(z',t) \,dz' d\lambda .
 \label{eq:evolut-f-omega}
\ee
In the next section we will show that this equation is basically identical to the quantum evolution equation of the Wigner function.

\section{Relationship to quantum mechanics}
Equation \eqref{eq:evolut-f-omega} was built purely on the two assumptions that the total probability and the logical entropy should be conserved in time. It is therefore striking that this equation bears a close resemblance to the evolution equation for the Wigner function $w$ in quantum mechanics \cite{Wigner1932,Hillery1984}, as will be discussed shortly.

The Wigner formalism is a representation of quantum mechanics in the classical position-momentum phase space $(x,p)$, which is strictly equivalent to the more usual Schr\"odinger or Heisenberg pictures. The state of a quantum system, either pure or mixed, is defined by a real function $w(x,p,t)$. The Wigner function is constructed  from the wave function for a pure quantum state or from the density matrix for a mixed state.
The Wigner function possesses many of the properties of standard probability distributions. For instance, it can be used to compute the average of a phase-space variable $A(x,p)$ as: $\langle A \rangle = \int \int w(x,p) A(x,p) dx dp $, where we have assumed the normalization $\int \int w(x,p) dx dp=1$. However, $w$ can take negative values, which precludes the possibility of interpreting it as a true probability density.

The Wigner function evolves in time according to an integro-differential equation that reads as:
\be
\frac{\partial w}{\partial t} + \frac{p}{m} \frac{\partial w}{\partial x} =
\frac{2\pi i}{h^2}\int \int \left[ V\left(x+{\lambda \over 2}\right) - V\left(x-{\lambda \over 2}\right) \right] \exp\left(\frac{2 \pi i (p-p') \lambda}{h}\right) \, w(x,p',t)\, dp' d \lambda , \label{eq:wigner-evolution}
\ee
where $V(x)$ is the potential energy.
Interestingly, the above evolution equation preserves in time both $\int \int w dx dp$ and $\int \int w^2 dx dp$, {\em but not higher powers of $w$}. This fact has motivated choosing the logical entropy as the natural definition of entropy in Wigner's quantum mechanics \cite{Manfredi2000}.

Now, we consider a Wigner function concentrated near a position $x=a$ and write: $w(x,p,t) = {\overline w(p,t)} \delta(x-a)$, where $\delta$ is the Dirac delta function. We also define $\Omega(x) \equiv 2\pi V(x)/h$, which has the dimensions of an inverse time.
Substituting into Eq. \eqref{eq:wigner-evolution} and integrating over $x$  yields
\be
\frac{\partial {\overline w}}{\partial t}  =
\frac{i}{h}\int \int \left[ \Omega\left(a+{\lambda \over 2}\right) - \Omega\left(a-{\lambda \over 2}\right) \right]  \exp\left(\frac{2 \pi i (p-p') \lambda}{h}\right)  {\overline w}(p',t)\, dp' d \lambda , \label{eq:wigner-evolution-delta}
\ee
which is identical to Eq. \eqref{eq:evolut-f-omega} with the correspondence $z\leftrightarrow p$.

It is quite remarkable that, based on the sole assumption that the probability density $f(z,t)$ preserves the total probability and the information (or entropy), we were able to construct an evolution equation \eqref{eq:evolut-f-omega} that is identical to the evolution equation of the Wigner function. In other words, the quantum evolution appears to stem uniquely from the property of conservation of the logical entropy (apart from the trivial conservation of total probability). This fundamental role played by the quantity $\int \int w^2 dx dp$ had already been noticed in earlier works \cite{Baker1958,Manfredi2000}

An important caveat is that the probability density $f(z,t)$ depends only the only variable $z$ (plus time), whereas the Wigner function depends on the two phase-space variables $x$ and $p$. For that reason,
we had to consider a Wigner function that is localized in space (Dirac delta function)
in order to establish the equivalence with the Wigner evolution equation. This is a significant difference, because it means overlooking a crucial feature of quantum physics, namely the existence of conjugate variables like position and momentum, whose simultaneous measurement is forbidden by the Heisenberg uncertainty principle.

In order to recover the full Wigner equation, we should work with probability distributions which, in the finite-dimensional case, depend on {\em two} indexes, such as $p_{ij}$, i.e., a matrix or tensor. The appropriate norm here appears to be the Frobenius norm $\|p\| = \sqrt{\sum_{i,j} p_{ij}^2}$, with the information defined as $I=\|p\|^2$. Then, in order to establish an evolution equation that preserves the norm, one would need to define a rotation of the tensor $p_{ij}$ in the appropriate space.
The generalization to an infinite dimensional space should lead to an evolution equation for a two-variable probability density $f(z_1, z_2,t)$, which will have to be compared to the full Wigner equation \eqref{eq:wigner-evolution} for $w(x,p,t)$. This extension is left for future work.

\section{Conclusions}\label{sec:conclusion}
In this work, we made use of the definition of logical entropy and information to extend the notion of probability to negative values.
Although negative probabilities have been considered extensively in the past (and often dismissed as unphysical), we argued that they fit nicely within the framework of the logical entropy. Indeed, rejecting negative probabilities would appear as rather arbitrary and odd if one trusts the definition of logical entropy.

Our strategy was to posit that all normalized probability distributions $\{p_i\}$ for which the logical entropy lies in the interval $[0, 1]$ are allowed, irrespective of the sign of the $p_i$s. Of course, the constraint on the entropy limits the absolute negative values that can be taken by the probabilities.

We also pointed out that the logical information has a straightforward interpretation as the square of the Euclidean norm of the probability vector in ${\mathbb R}^n$, or the $L^2$ norm in the case of a continuous probability density. This simple geometric property is extremely fruitful to derive various interesting properties. In particular, the set of allowed probability distributions may be seen as the intersection of a hypersphere and a hyperplane in ${\mathbb R}^n$.

In order for the total probability and entropy to be conserved in time, the probability vector must {\em rotate} in the appropriate space, and this rotation is defined by an antisymmetric matrix. We next generalized this rotation to the infinite dimensional case (continuum). Quite remarkably, this leads to an evolution equation for the probability density $f(z,t)$ that is virtually identical to the Wigner equation for a quantum system, at least when one considers only the momentum variable.
These findings highlight the fundamental role played by the logical entropy in the mathematical structure of quantum mechanics.

Our future program is to prove that the full Wigner formulation of nonrelativistic quantum mechanics may be deduced from just two simple postulates: (i) conservation of the total probability $\int \int w(x,p,t) dx dp$ and (ii) conservation of the  logical information $h \int \int w^2(x,p,t) dx dp$. For this, one should extend the present derivation to probability densities that depend on two variables, namely position and momentum.
Once realized, this program would establish an alternative axiomatic foundation to nonrelativistic quantum mechanics.

\bigskip

\noindent{\bf  Acknowledgments}\\
I wish to thank David Ellerman for his thorough reading of a draft of this paper and several insightful comments.

\bibliography{entropy}

\end{document}